\documentclass[aps,prstab,preprint,showpacs,groupedaddress]{revtex4}

\usepackage{graphics,epsfig}
\begin{document}
\title{Evolutionary Minority Game with Multiple Options}

\author{Hong-Jun Quan$^{1,2}$, P. M. Hui$^{2}$, C. Xu$^{3}$, K. F Yip$^{2}$}

\address{$^{1}$ Department of  Physics, South China University of
Technology, Guangzhou  510641, China\\
$^{2}$ Department of Physics, The Chinese University of Hong Kong,\\
Shatin, New Territories, Hong Kong \\
$^{3}$ Department of  Physics, Suzhou University, Suzhou 215006,
China}


\begin{abstract}

We propose and study an evolutionary minority game (EMG) in which
the agents are allowed to choose among three possible options.
Unlike the original EMG where the agents either win or lose one
unit of wealth, the present model assigns one unit of wealth to
the winners in the least popular option, deducts one unit from the
losers in the most popular option, and awards $R$ ($-1<R< 1$)
units for those in the third option. Decisions are made based on
the information in the most recent outcomes and on the
characteristic probabilities of an agent to follow the predictions
based on recent outcomes. Depending on $R$, the population shows a
transition from self-segregation in difficult situations
($R<R_{c}$) in which the agents tend to follow extreme action to
cautious or less decisive action for $R>R_{c}$, where $R_{c}(N)$
is a critical value for optimal performance of the system that
drops to zero as the number of agents $N$ increases.

\vspace*{0.1 true in}



\end{abstract}
\maketitle


\newpage
\section{Introduction}
Agent-based models of complex adaptive systems \cite{Holland} have
recently attracted much attention among scientists in different
research areas \cite{general1,book,general2}. Typically, these
models consist of a competing population in which agents decide
based on some global information. The information is, in turn, the
result of the collective behavior of the population. An
outstanding example of an agent-based model is that of the
Minority Game (MG), which is a binary version of Arthur's bar
attendance problem \cite{arthur}, proposed by Challet and Zhang
\cite{Challet1}. In the MG, agents compete to be in a minority
group and make decisions based on information generated by the
actions of the agents in previous rounds without direct
interaction among the agents.  The population show interesting
cooperative actions \cite{savit,Johnson0,hart,marsili}.  The MG
forms the basis of many interesting agent-based models in recent
years \cite{Challet3}, and many of its variations were proposed in
connection to possible applications in areas such as complex
systems and econophysics \cite{book}. These variations differ from
the original MG mainly in the way in which the adaptive ability of
the agents is introduced into the models.

In the Evolutionary Minority Game (EMG) proposed by Johnson {\em
et al.} \cite{Johnson1,Lo1,Hui,Lo2,freezing}, an odd number $N$ of
agents decide to choose one of two options, 0 or 1, at each time
step. The agents who are in the minority (majority) group win
(lose) and are awarded (deducted) one point. Every agent holds the
same dynamical strategy of simply following the most recent trend,
together with two individual parameters, namely a score and a
probability $p$ ($0 \leq p \leq 1$). The probability $p$ is the
chance that an agent decides to follow the strategy's prediction
and $1-p$ is the chance that the agent decides to act opposite to
the current trend.  The $p$-value of an agent is allowed to
change, if the agent does not perform well.  If the score of an
agent drops below a threshold $d$ ($d \leq 0$), the agent replaces
his $p$-value by a new value randomly taken within a range $r$ of
the original $p$-value and his score is reset to zero. The most
interesting feature in EMG is that agents who behave in an extreme
way (i.e., using $p \approx 0$ and $p \approx 1$) perform better
than the cautious ones (i.e., using $p \approx 0.5$)
\cite{Johnson1,Lo1}. This in turn leads to a self-segregation of
the population in the sense that the distribution of $p$-values
tends to peak at $p \approx 0$ and $p \approx 1$.  Hod and Nakar
\cite{Hod} studied a modified version of EMG with a biased payoff
function, i.e., the points awarded to winners and deducted from
losers are different. The authors found that self segregation
occurs only if the ratio $\overline{R}$ of the point awarded to
the point deducted is greater or equal to unity.  For
$\overline{R}<1$, i.e., corresponding to difficult situations in
which winning in a turn cannot compensate losing in another turn,
the distribution of $p$-values shows the feature of clustering by
which agents tend to be more cautious or less decisive and take on
$p$-values around $0.5$.

Motivated by real-life scenarios in which there may be more than
one winning decisions with different payoffs and by the different
behavior of the modified EMG \cite{Hod} for $\overline{R} \geq 1$
and $\overline{R} <1$, we propose and study a generalized version
of EMG with three possible options. The agents in the most (least)
popular option are deducted (awarded) one point, while the agents
in the third option are awarded $R$ points, where $R$ may be
positive or negative with $|R| < 1$. Each agent has the
probabilities $(p_{1},p_{2},p_{3}= 1-p_{1}-p_{2})$ to follow the
prediction of the most current trend of the outcomes. It is found
that a transition from self-segregation in the distribution of
$p$-values to cautious behavior occurs as $R$ changes. At some
critical value $R_{c}$, the system performs optimally. The plan of
the paper is as follows. In Section II, the three-option EMG is
defined. We present and discuss results of detailed numerical
simulations in Section III. Results are summarized in Section IV.

\section{Three-option EMG}

The present model represents a generalization of the original EMG
\cite{Johnson1} from two options to three options. Our model is
motivated by situations in which there may be more than one
winning or losing options. Many agent-based models were proposed,
for example, with possible applications related to the features
observed in financial markets \cite{book}. Taking the financial
market as an example, agents may have many choices on whether to
invest on the stock market, the money market, bonds, derivatives,
etc., depending on the availability of funds, the level of risk
that one may want to handle, and the agent's confidence level.
Usually a market of higher risk also brings a higher earning if
the correct investment is made. Within a chosen market, there are
good investments and bad investments. Thus there may be several
good decisions among the various markets at the same time. A
similar situation also happens when one decides on the options for
pension fund investments, for which choices such as growth,
balanced, and stable funds are available. On the more entertaining
side, betting on horse racing also provides choices of different
levels of risk through different betting pools like win, place,
quinella (i.e., betting on the first and second places not in
order), etc. All these situations involve the choice of several
options, with a different payoff for each option.

To include multiple options into EMG, we consider a number $N$ of
agents where $N$ is not a multiple of $3$. Each agent chooses
among three options, say, 1,2 and 3, at each time step. The agents
in the side with the least (most) number of agents, i.e., in the
minority (majority) side, win (lose) and are awarded (deducted)
one point, while the agents in the third option with intermediate
number of agents get $R$ point, where $R$ may be negative or
positive and $|R| <1$. The outcome in increasing number of agents,
i.e., increasing popularity, choosing the three options
$(\Omega_{1}(t),\Omega_{2}(t),\Omega_{3}(t))$ at the time step $t$
forms the publicly known information of the game. For example, the
outcome $(3,1,2)$ implies that option 3 (2) is the least (most)
popular option.  Every agent uses the set of outcomes of the most
recent $m$ time steps,
$\mu(t)=\{(\Omega_1(t-m+1),\Omega_2(t-m+1),\Omega_{3}(t-m+1)),
\cdots, (\Omega_1(t-1),\Omega_2(t-1),\Omega_{3}(t-1))\}$, as the
information to predict the current trend.  For given $m$, there
are a total of $3^{m} \times 2^{m}$ different $\mu(t)$.  The
agents are also provided with a strategy that corresponds to the
outcomes of the most recent occurrences of each of the $3^{m}
\times 2^{m}$ possible $\mu$, as in the original EMG
\cite{Johnson1}. This strategy gives the most recent trend, i.e.,
what happened in terms of popularity among the three options the
last time that a particular $\mu$ occurred.  The strategy is
dynamical in the sense that it changes as the game proceeds.

At any time step of the game, each agent carries his own set of
probabilities $(p_{1},p_{2},p_{3}=1-(p_{1}+p_{2}))$, with $0 \leq
p_1 \leq 1$, $0 \leq p_2\leq 1$, and $0 \leq (p_1+p_2) \leq 1$.
Given the most recent $m$ outcomes, i.e., for given $\mu(t)$, an
agent has the probability $p_{1}$ ($p_{3}$) to choose the
predicted least (most) popular option as suggested by the
strategy, and probability $p_{2}$ to follow the prediction on the
intermediate option.  Initially, the $p$-value of each agent is
assigned randomly, and the score is set to zero.  As the game
proceeds, the performance is registered in the score of the
agents. If an agent's score falls below a certain threshold value
$d$ ($d \leq 0$), he is allowed to replace his $p$-values by new
values of $p_1$ and $p_2$ within a range $r$ centered at the
current values $p_1$ and $p_2$, and his score is reset to zero.
Since $p_{3} = 1-p_{1}-p_{2}$, it is sufficient to work in the
$p_{1}$-$p_{2}$ space. A reflective boundary condition
\cite{Johnson1} is imposed in the ($p_1,p_2$)-space to ensure that
the requirements $0 \leq p_i\leq 1$ ($i = 1,2,3$) are satisfied.
Therefore, evolution comes in by allowing agents to modify their
$p$-values. The present model is different from an earlier version
of multiple-choice EMG proposed by Metzler and Horn \cite{Metzler}
in that we allow for possibly more than one winning options. The
model is also different from the previously proposed
multiple-choice MG \cite{rodgers,chow1} in that the second least
popular option may also be a winning choice and adaptability is
introduced through the $p$-values instead of different strategies
assigned to the agents.

\section{Results}

Extensive numerical simulations have been carried out to study the
$p$-value distribution $P(p_{1},p_{2})$ in the population.
Consider a system with $N=1001$ agents, $m=3$, $r=0.2$ and $d=-4$.
Figure 1 shows typical $P(p_1,p_2)$ on a grey-scale 2D plot
projected on to the $p_{1}$-$p_{2}$ plane for three different
values of $R$ ($R$=-0.5, 0.043, 0.5). Results are obtained by
averaging over 10 independent runs for each value of $R$, with an
initially uniform distribution. In each run, the distribution is
obtained in a time window of $10^5$ time steps, after transient
behavior dies off. In constructing $P(p_1,p_2)$, the $p_1$-axis
and $p_2$-axis are each divided into 25 divisions, i.e., each
division corresponds to 0.04. The distribution $P(p_{1},p_{2})$ is
normalized so that $\int \int P(p_1,p_2)dp_1dp_2=1$. The
distribution $P(p_1,p_2)$ for $R$=-0.5 is rather flat, implying
that there is no advantage in having any specific set of
($p_1,p_2,p_3$) over others. As $R$ increases, the number of
agents taking on extreme actions (($p_1,p_2,p_3$) $\approx
$(1,0,0) or (0,1,0) or (0,0,1)), i.e., persistently making a
certain choice, increases, and there exists a small range of $R$
in which $P(p_1,p_2)\approx $ $0$ for intermediate values of
$(p_1,p_2)$ (see Fig.1(b)). The distribution $P(p_1,p_2)$ is
nearly symmetric about ($p_1\approx 0.3,p_2\approx 0.3$) with
peaks around $(p_1,p_2)$=(0,1),(1,0) and (0,0). This behavior is
analogous to the segregation of agents into extreme actions in the
original EMG \cite{Johnson1,Lo1,Hui,Lo2}. The result implies that
in order to flourish in such a population in difficult situations
($R<0$), an agent should behave in an extreme way. This
segregation behavior leads to an enhancement in the performance of
the population as a whole in that the number of agents in the
least popular option takes on a value approaching the limit $N/3$,
as allowed by the definition of the winning minority side.
Although results are shown only for the case of $m=3$ and $d=-4$,
the steady state distribution $P(p_1,p_2)$ does not depend
sensitively on $m$ and $d$, and the initial distribution. When $R$
is further increased (see Fig.1(c)), the game allows for more
winners than losers.  In this case, the agents become less
decisive or cautious. The distribution shows a peak at about
($p_1\approx 0.36,p_2\approx 0.28$), implying that the strategy's
prediction of the winning option will be too crowded to win. It
should be noted that the steady state distribution in this case is
dependent on the initial distribution. It is analogous to the
freezing phenomena in the original EMG when the resource level is
high \cite{freezing}. Here $R$ plays the role of a resource level
in that a positive $R$ implies a majority of agents will earn a
reward per turn. The lifespan $L(p_1,p_2)$, defined as the average
duration for an agent holding ($p_1,p_2$) between modifications of
$p$-values, shows similar behavior as the distribution
$P(p_1,p_2)$.

The performance of the system can be related to the average number
of agents in each option and the fluctuations (variance) in the
number of agents making a particular choice over time. Figure 2
shows the average number of agents (right axis) in the most
(least) popular option and the third option and the variance
$\sigma ^2/N$ in the number of agents (left axis) for different
values of $R$ ($m$=3, $d$=-2 and $r$=0.2), respectively. The
average number of agents in each option takes on values close to
$N/3$, with the winning option determined by a margin of about 10
agents.  For $R<0$, i.e., when there are more losers than winners,
the number of agents in each option is insensitive to $R$. As $R$
increases to the vicinity of $R \approx 0$, the average number of
agents in the least (most) popular option also reaches a maximum
(minimum) at some value $R_c$ with $R_{c} \approx 0$, signifying
an enhanced performance of the system. Interesting, the number of
agents in the third (intermediate) option remains flat and close
to $N/3$ over a wide range of $R$. As $R$ is further increased
($0<R<1$), more agents choose to follow the strategy's prediction
on the least popular option and the predicted option becomes the
most popular and hence too crowded to win. As a result, fewer
agents choose the least popular (winning) option.

The dependence of the variance $\sigma ^2/N$ on $R$, for each of
the options, is non-monotonic and shows a corresponding drop and
reaches a minimum at $R \approx R_{c}$. A smaller fluctuation
implies a higher number of winners per turn, and hence better
performance as a whole. The results imply an optimum cooperation
in the population.  For $R > R_{c}$, the variance increases with
$R$. For comparison, the dashed line gives the the variance
$\sigma_{rand}^2/N$ for random decisions in a multiple-choice game
which is given by
\begin{equation}
\frac{\sigma_{rand}^2}N=\frac
1N\sum_{m=0}^Nq^{N-m}(1-q)^mC_N^m((N-m)-Nq)^2
\end{equation}
where $q$ is the probability of taking a side. For three
three-option case, $q=1/3$, and $\sigma_{rand}^2/N=2/9$. It should
be noted that the variance $\sigma ^2/N$ for all the values of $R$
studied is smaller than $\sigma_{rand}^{2}/N$, implying that the
agents are actually benefitted from the evolutionary nature of the
game.  Thus, $R_{c}$ signifies an {\em optimal} reward (resource)
of the system in that at $R=R_{c}$, a {\em maximum fraction} of
the total possible reward per turn, e.g., $(N-2)/3 - (1-R)(1 +
(N-2)/3)$ for $N=1001$, is actually awarded to the agents.

To look closer into the change in the agents' behavior for
different values of $R$, it is convenient to study the separation
in the $p$-values among the agents in the population.  We define
an averaged separation $S$ by
\begin{equation}
S =\frac {2}{N(N-1)T}\sum_{t=1}^T \sum_{i,j=1}^N
\sum_{\alpha=1,2,3} [p_{\alpha}(i) - p_{\alpha}(j)]^{2},
\end{equation}
where $T$ is the time window for taking the average.  The
separation $S$ is thus a mean over the difference squared of the
$p$-values of the agents.  Segregation in $p$-values among the
agents corresponds to a larger value of $S$. Figure 3 shows the
average separation $S$ for different population sizes $N$=1001,
500, 200 and 101 ($m=3,r=0.2,d=-4$). For negative $R$, the
fraction of agents taking extreme action is small, so the average
reduced distance is small. As $R $ increases, more agents show
extreme behavior, resulting in a rapid increase in $S$ with a
maximum near $R = R_{c}$. The peak in $S$ becomes narrower as the
system size becomes larger. For $R > R_{c}$, the separation $S$
drops quite rapidly and attains a value smaller than that of $R <
0$. The results in Fig.3 indicate that we may identify the value
of $R_{c}(N)$ as the value of $R$ at which the separation $S$
attains a maximum. The resulting $R_{c}(N)$ (inset of Fig.3) shows
that $R_{c}$ drops monotonically with $N$ to a value of $R_{c}
\approx 0.04$ for $N=1001$.  The trend of the results proposes
that $R_{c}$ may approach the value of $0$ as $N$ is further
increased.

\section{Summary}

We have proposed and studied a three-option evolutionary minority
game with different payoffs to the agents in each of the options.
Our model is a generalization of the original two-option EMG.
Interesting behavior, including self-segregation into extreme
groups and non-decisiveness, are found, depending on the payoff to
the third option (besides the minority and majority options). In
difficult situations ($R<R_c$), segregation occurs and the agents
prefer to take extreme action (($p_1,p_2,p_3$) $\approx $ (1,0,0)
or (0,1,0) or (0,0,1)). For $R>R_c$, the $p$-values of the agents
tend to cluster at a common location.  Detailed analysis on the
average number of agents taking each option and the corresponding
variance revealed an optimal performance of the system at some
value $R_c$. Studies on systems of different sizes showed that
$R_{c}$ drop nearly to zero as the number of agents $N$ in the
system increases. A negative and positive value of $R$ correspond
to very different situations in that for $R<0$ there is a majority
of losers in the population, while for $R>0$ the system allows for
a majority of winners.

\acknowledgments{This work was supported in part by the Research
Grants Council of the Hong Kong SAR Government through the grant
CUHK 4241/01P.}

\newpage
\centerline{\bf FIGURE CAPTIONS}

\bigskip

\noindent Figure 1. The distribution $P(p1,p2)$ of the $p$-values
among the agents in a system of $N=1001$ with $r=0.2$, $m=3$,
$d=-4$ for (a) $R=-0.5$, (b) $0.043$, and (c) $0.5$, projected on
to the $p_{1}$-$p_{2}$ plane. The grey-scale indicates the value
of $P(p_{1},p_{2})$.

\bigskip

\noindent Figure 2. The average number of agents (right axis)
taking the three options (different symbols) and the corresponding
variances $\sigma ^{2}/N $ (left axis) for different values of
$R$.  Other parameters are $m$=3, $d$=-2 and $r$=0.2.  The lines
are guide to eye.  The dashed line gives the variance in Eq.(1)
corresponding to random decisions.

\bigskip

\noindent Figure 3. The average separation of probabilities $S$
among the agents for different number of agents $N$=1001, 500, 200
and 101 in the system.  Other parameters are $m=3$, $r=0.2$,
$d=-4$. The separation $S$ shows a maximum at $R_{c}(N)$.  The
inset shows the dependence of $R_{c}$ on $N$.

\newpage
\begin{figure}
\epsfig{figure=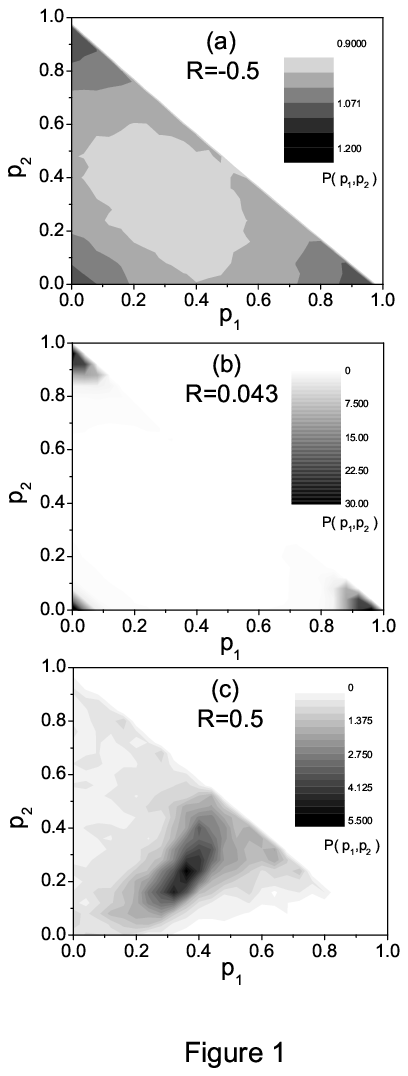,width=10.5cm}
\label{figure1}
\end{figure}

\newpage
\begin{figure}
\epsfig{figure=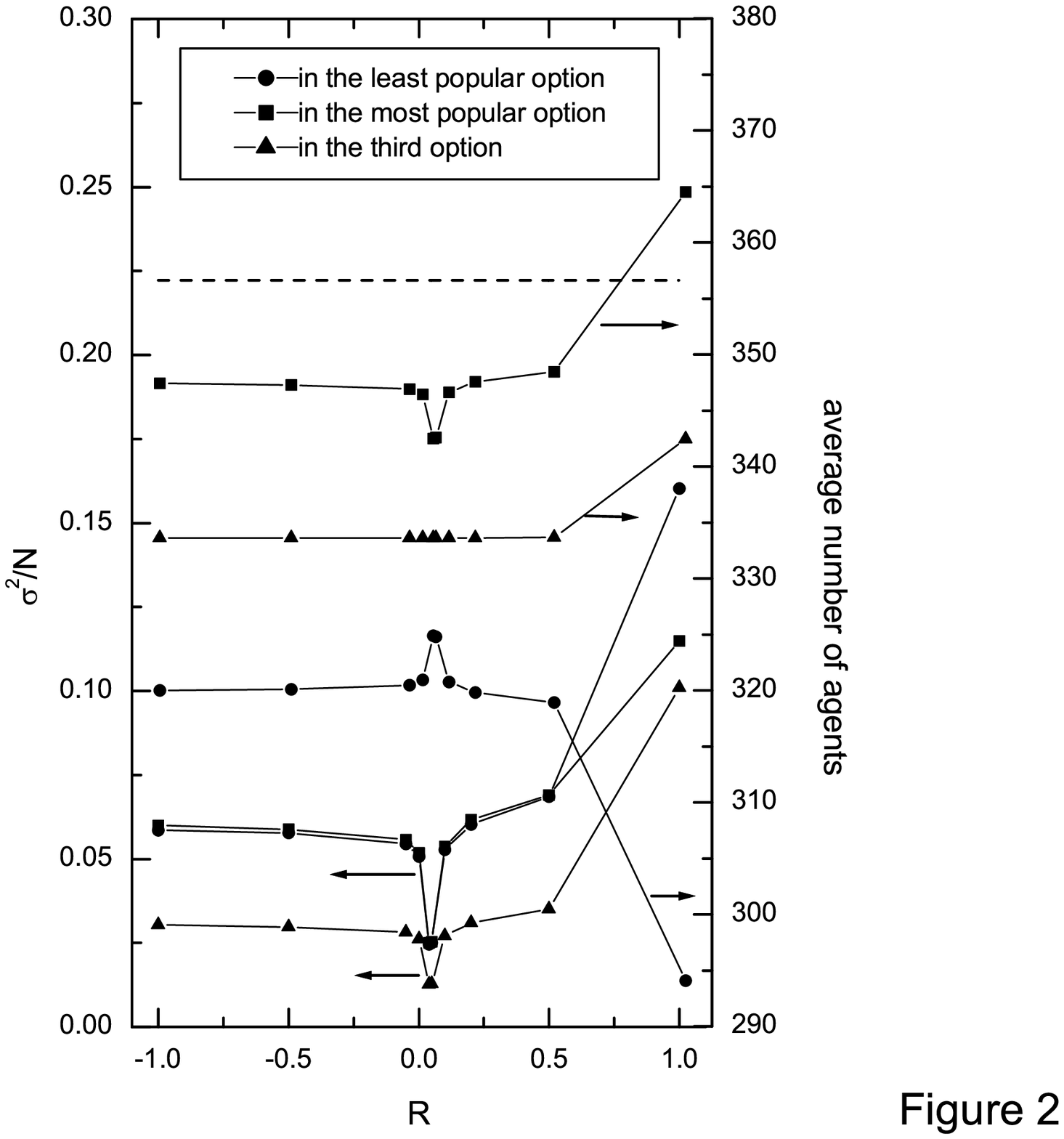,width=\linewidth} \label{figure2}
\end{figure}

\newpage
\begin{figure}
\epsfig{figure=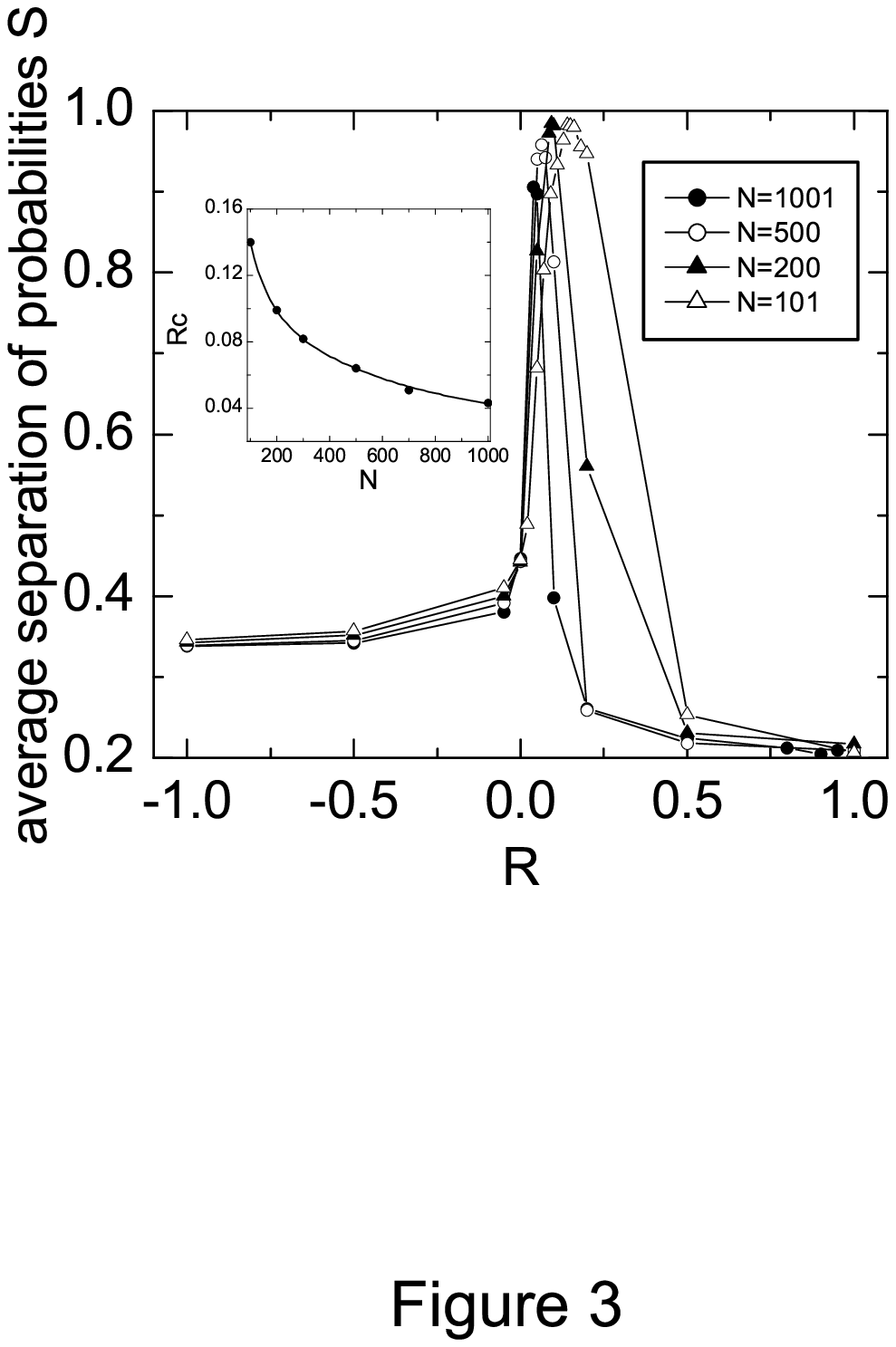,width=\linewidth} \label{figure3}
\end{figure}



\end{document}